\documentstyle[aps,prl,epsf,floats]{revtex}
\newcommand{\beq}{\begin{equation}}
\newcommand{\eeq}{\end{equation}}
\newcommand{\ds}{\displaystyle}
\newcommand{\beqar}{\begin{eqnarray}}
\newcommand{\eeqar}{\end{eqnarray}}

\begin{document}
\draft

\twocolumn[\hsize\textwidth\columnwidth\hsize\csname@twocolumnfalse%
\endcsname

\title  {
 Irreversibility, steady state, and non-equilibrium physics in 
 relativistic heavy ion collisions
         }
\author {
E.~E.~Zabrodin,$^{1,2}$ 
L.~V.~Bravina,$^{1,2}$
H.~St{\"o}cker,$^{1}$ and W.~Greiner$^{1}$ \\
  }
\address{$^1$
 Institute for Theoretical Physics, University of Frankfurt,
 Robert-Mayer-Str. 8-10, D-60054 Frankfurt, Germany 
         }
\address{$^2$ 
 Institute for Nuclear Physics, Moscow State University,
 119899 Moscow, Russia
         }
\date {January 20, 1999}
\maketitle

\begin{abstract}
Heavy ion collisions at ultrarelativistic energies offer the 
opportunity to study the irreversibility of multiparticle processes. 
Together with the many-body decays of resonances, the multiparticle
processes cause the system to evolve according to Prigogine's steady 
states rather than towards statistical equilibrium. 
These results are general and can be easily checked by any 
microscopic string-, transport-, or cascade model for heavy ion 
collisions. The absence of pure equilibrium states sheds light on 
the difficulties of thermal models in describing the yields and 
spectra of hadrons, especially mesons, in heavy ion collisions at 
bombarding energies above 10 GeV/nucleon. 
\end{abstract}
]


The hypothesis that local equilibrium (LE) is attained by the system 
of two heavy ions colliding at ultra-relativistic energies is a basic
assumption of macroscopic thermal- and hydrodynamical models of heavy
ion collisions. The idea was pushed by Fermi \cite{Fer50} and Landau
\cite{Land53,LaBe56} almost 50 years ago for hadron-hadron collisions.
Despite the long history of theoretical and experimental attempts 
there is no rigorous proof of LE yet. 
The present paper shows that the irreversibility of multiparticle
processes, proceeding e.g. via string decays, causes these systems to 
evolve according to Prigogine's \cite{GlPr71} steady state solution, 
rather than towards statistical equilibrium.

Using Bogolyubov's hierarchy of relaxation times in 
non-equilibrium statistical mechanics \cite{Bogo46} one usually
considers the following scheme: Suppose that in the initial stage the 
system is far from equilibrium. To describe it one has to introduce a 
set of various many-particle distribution functions rapidly varying in 
time. Then, due to interactions between the particles, correlations of 
the distribution functions occur within very short time intervals
which are typically on the order of the collision time. 

This is the 
kinetic stage -- all many-particle distribution functions may be 
derived from the single one-particle distribution function. For times
significantly larger than the collision time the number of parameters
characterizing the system is reduced further to very few average
values, namely the number of particles, their energy and velocity, 
i.e. to the moments of the distribution function. At this stage the 
system behavior is governed by hydrodynamics.

Unlike in non-relativistic mechanics, in relativistic heavy ion 
collisions the relaxation picture is more complex because of
multiparticle processes. Here the number of particles and their 
composition are not conserved. Newly produced particles are not
thermalized (even if they appear to be, see \cite{SBM,Becc96}) 
and this circumstance causes a delay in achieving
equilibration. The equilibration time may appear too long as compared
to the typical lifetime of the expanding system. Due to the lack of a
rigorous first-principles theory of nuclear reactions at relativistic 
energies, the approach to LE is investigated mainly by virtue of 
dynamical calculations provided by microscopic semiclassical Monte 
Carlo models 
\cite{Geig95,Bere92,Faes94,QGSM,RQMD,UrQMD,LV98plb,LV98jpg} 
which have been intensively studied during the last 15 years.

The analysis of the space time evolution picture
obtained in these models reveals that the whole system of colliding 
nuclei never attains a global equilibrium state after the initial 
non-equilibrium stage.
Still, there is, in principle, a possibility of the occurrence of 
local equilibrium (e.g. in the central cell), because the approach
to LE does not depend on the assumptions of the
presence of a heat reservoir, of Gibbs ensembles, etc. 

Our study has been inspired by the finding that quasi-stable
states are present in partonic and hadronic matter, as observed 
independently in dynamical simulations \cite{Geig95,Belk98,Bran99}. 
On the partonic level an analysis of the thermalization of partons 
has been performed by the late Klaus Kinder-Geiger \cite{Geig95}.
Equilibration of hadronic matter has been studied, e.g., in the 
Quantum Molecular Dynamics models 
\cite{Bere92,Faes94,UrQMD,LV98plb,LV98jpg,SHSX98}. 
These simulations have shown that at high energies neither 
the global system nor its central part seem to reach the state
of chemical equilibrium (in the sense of statistical mechanics)
\cite{Geig95,LV98plb}. This observed feature is not solely 
restricted to microscopic models. 
To describe, for instance, the experimental data on yields
of strange particles in heavy ion collisions at 200 AGeV \cite{wa85}
or hadron multiplicities at 158 AGeV \cite{na49}
the standard statistical model of the ideal hadron gas has been 
modified to invoke the hypothesis of chemical non-equilibrium 
\cite{LeRa99,YeGo98} as well.

Does this simply imply that the hadronization time is shorter than 
the equilibration time? - Not necessarily! In the present paper we 
show that dissipative processes, such as multiparticle production 
via strings and many-body ($N \geq 3$) decays of resonances, 
dominating at high energies, can lead to the creation of a 
stationary state called steady-state. This steady-state does 
{\bf not} coincide with a pure ``conventional" equilibrium state, 
as assumed in the statistical models.  

Consider first the necessary and sufficient criteria of LE in the
central zone of nuclear reactions, which is usually 
analyzed in microscopic calculations:\\
{\bf Necessary conditions}: {\bf (i)} absence of significant flow 
effect in the central cell; isotropy of the velocity distributions, 
and {\bf (ii)} isotropy of the diagonal components of the pressure 
tensor,
\beq
P_x = P_y = P_z = {1 \over {3V}}\, \sum_i
\frac{p^2_{i\{x,y,z\}}}{(m_i^2~+~p_i^2)^{1/2}}~.
\label{eq1}
\eeq
Here $V$ is the volume of the cell and $m_i,p_i$ are the mass of 
the $i$-th hadron and its momentum, respectively.\\ 
{\bf Sufficient conditions}:
{\bf (iii)} thermal equilibration
which manifests itself in the 
time independence of the hadronic spectra after a certain period, 
and
{\bf (iv)} chemical equilibration, i.e. the time independence of 
different hadronic yields.

The necessary conditions look quite simple and evident: Local 
equilibrium may not be reached in symmetric nuclear collisions earlier 
than for the time $t^{pass}=2R/(\gamma_{cm} v_{cm})$, during which 
noninteracting Lorentz contracted nuclei of radius $R/\gamma_{cm}$, 
which stream freely with the velocity $v_{cm}$, would
have passed through each other. Apparently, early in the collision
this is the origin of a substantial initial longitudinal collective 
flow of hadrons in the cell, which distorts the equilibration 
picture at the very beginning of the reaction. After $t^{pass}$ this 
non-equilibrium component rapidly drops \cite{LV98plb}. 
In \cite{LV98jpg} it has been reported that a stage of {\it kinetic\/} 
equilibrium is attained in heavy ion collisions in a central cell of 
volume $V = 5\times 5\times 5 = 125$ fm$^3$ at about $t \cong 10$ 
fm/$c$, irrespective of the energy of the colliding nuclei from 10.7 
AGeV (AGS) to 160 AGeV (SPS). Isotropy of both the pressure and the 
velocity distributions of hadrons characterizes, {\bf without} the
sufficient conditions {\bf (iii)} and {\bf (iv)},
however, a pre-equilibrium stage of the reaction rather than an 
equilibrium one! In a fully equilibrated system conditions
{\bf (iii)-(iv)} must be satisfied as well.

This is the crucial point in our discussion: The statistical 
thermodynamics of many-particle systems \cite{LaLi80,deGr80} 
determines the thermal equilibrium as the state with maximum entropy. 
Once thermal equilibrium is attained, the velocity distributions of 
different particles must be isotropic. If the total number of 
particles is conserved, {\it kinetic\/} equilibrium is equivalent to 
{\it thermal\/} equilibrium \cite{deGr80}. But: this 
equivalence is broken, both in chemical reactions and in high energy 
physics. 

Indeed, if the mixture of reacting substances is in the ``true"
equilibrium, then the rates of each chemical reaction must be the
same for the direct and inverse processes \cite{GlPr71,LaLi80}. 
However, in a cyclic process, in which the concentrations of the
reacting substances are time independent, but the partial reaction 
rates $\omega_j = \omega_j^{\rm dir} - \omega_j^{\rm inv}$ are 
non-zero, the system is in a stationary state, which may be far from 
the equilibrium \cite{GlPr71}.

Consider now an ideal thermostat which contains a few thousand 
protons with an energy $E \gg m_p c^2$, where $m_p$ is the mass of 
proton and $c$ is the light velocity. For the sake of simplicity we 
exclude the (slow) weak processes from this scenario, focusing on 
strong interactions only. Then, even if the initial momentum 
distribution of the protons is Maxwellian, thermal equilibrium (in 
the sense of a state of maximum entropy) is not reached yet. Many new 
particles, mostly pions, will be produced as a result of initial 
proton-proton and, later, proton-pion, etc. collisions. 
When the system will finally reach equilibrium, it will consist of a 
large number of pions (and heavier mesons) with an admixture of 
baryons (and antibaryons) whose net number is conserved. The final 
temperature must, of course, be much lower than the initial one --
kinetic energy has been transformed into mass (of produced 
particles). But: will the particle abundances be the same as those 
given by the statistical mechanics of an ideal hadron gas?
In other words, will the final state be the state of thermal and 
chemical equilibrium, in which any direct and inverse hadronic 
processes will be taking place on average at the same rate?  

This problem is closely related to the principle of detailed balance 
and to the irreversibility of multiparticle processes. 
To avoid ambiguities, we would like to stress that the definition of
detailed balance in quantum mechanics (DB$^{QM}$) does not coincide
with the definition of detailed balance in statistical physics
(DB$^{SP}$) and chemistry. Detailed balance in the sense of quantum 
mechanical invariance under time reversal implies that the transition 
amplitudes of the direct and the inverse processes must be of the 
same magnitude,
\beq
\ds
|M_{a\rightarrow b}| = |M_{a\leftarrow b}|\quad.
\label{eq2}
\eeq
In statistical physics and chemistry, the principle of detailed 
balance requires that (in thermostatic equilibrium of a system) every 
separate reaction between its components
is in itself in equilibrium, i.e. the rates of the direct and inverse 
processes are the same. To clarify the difference between DB$^{QM}$
and DB$^{SP}$, consider the process of multiparticle production, e.g.
in string excitation:
\beq
\ds
a + b \longrightarrow x_1 + x_2 + \ldots + x_n,\quad n \gg 1
\label{eq3}
\eeq
According to Fermi's Golden Rule, the probability of $n$ particle 
production reads
\beq
\ds
{\rm d} R_n = \frac{2 \pi}{\hbar} |M_{a+b\rightarrow n}|^2\, 
\prod_{i=1}^{n} 
{\rm d}^4 p_i \, \delta^4 \left( p - \sum_{i=1}^n p_i \right) \quad,
\label{eq4}
\eeq
where $p$ is the total four-momentum, $p_i$ is the four-momentum of
$i$-th particle, and $|M|$ is the amplitude of the process.
The last factor is the space factor, which is fully determined 
by the kinematics of the reaction \cite{ByKa73}. 
Although $|M_{a+b\rightarrow n}| = |M_{a+b\leftarrow n}|$ and,
therefore, DB$^{QM}$ is satisfied, the rates of the direct and the
inverse processes, $R_{a+b\rightarrow n}$ and $R_{a+b\leftarrow n}$, 
are different, due to different space factors. This means that 
DB$^{SP}$ is not fulfilled. 
Note that the principle of detailed balance in particle physics has
been verified for the reactions
\beqar
a + b & \longleftrightarrow & c + d\quad, \\
\label{eq5}
a + b & \longleftrightarrow & c\quad,
\label{eq6}
\eeqar
where $a,b,c,d$ denote hadrons and their resonances. These processes
are time reversible, because the space factors (or the densities of 
states) of the initial and final states are essentially the same. 

\begin{figure}
\centerline{\epsfysize=8.0cm \epsfbox{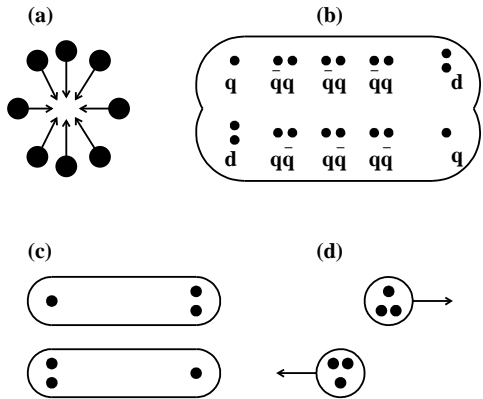}}
\caption{
Irreversible process: recombination of several hadrons into two
baryons. Schematic diagram of (a) many-particle collision of hadrons;
(b) rearrangement of their quarks and subsequent annihilation of
$q \bar q$-pairs; (c) production of two quark-diquark strings, which
shrink to (d) two baryons of high energy.
}
\label{fig1}
\end{figure}

The space factors are rapidly varying functions of $n$. Therefore, 
the matrix elements $|M|$ may be replaced by average values, and a 
situation typical for statistical mechanics is obtained: the 
probability of a state is proportional to the volume of the accessible 
phase space. In other words, multiparticle processes are irreversible 
in time, because they increase the local entropy of the system. 
Consequently, the processes of recombination of many hadrons into one
string, and two strings colliding to form a couple of ground state 
hadrons (Fig.~\ref{fig1}) of high energy are strongly suppressed, 
because they violate (locally) the Boltzmann $H$-theorem.

The irreversibility of multiparticle processes (e.g., strings which
provide a steady source of new particles) first drives and then keeps 
the hadronic system out of the total chemical equilibrium, i.e. out 
of the full detailed balance in the sense of statistical mechanics. 
On the other hand, the DB$^{SP}$ principle is the basic assumption of 
the statistical model (SM) of the ideal hadron gas 
\cite{SOG81,GMQY94,BrMSWX96,CEST97} 
and variations like the statistical bootstrap model (SBM) \cite{SBM}. 
Therefore, simply extracting the energy density $\varepsilon$, the
baryon density $\rho_{\rm B}$ and the strangeness density 
$\rho_{\rm S}$ of the system at a given time and inserting these 
values as an input into the statistical model will be giving 
misleading results until all multiparticle processes in the system
will have ceased. 

Still, the conditions {\bf (iii)-(iv)} may be fulfilled, even if the 
full detailed balance is not reached yet.
Such states, which may be stable or not, but are out of local 
equilibrium, have been dubbed steady states of the system 
\cite{GlPr71}.
To decide whether or not a steady state is attained in a 
microscopic model of heavy-ion collision, the system must be 
compared with the quasi-equilibrated (in the sense of the criteria
{\bf (i)-(iv)}) infinite matter, as simulated within the same 
microscopic model. In \cite{LV98jpg} it was shown that the yields
and energy spectra of hadrons in a central cell are -- after 
$t \cong 10\,$fm/$c$ -- very close to those values calculated for 
infinite hadron matter \cite{Belk98} with the same 
$\varepsilon$, $\rho_{\rm B}$ and $\rho_{\rm S}$.
This is a strong indication on the occurrence of a steady state. 

In conclusion, we have discussed the relaxation of hadronic matter 
produced in the central zone of heavy ion collisions in the energy 
range spanning from AGS to RHIC. Apparently, dissipative $N$-body 
($N \gg 1$) decays of strings and resonances, i.e. multiparticle 
processes, are irreversible in time: 
the probability of $N$ particles \\
1) to collide simultaneously in a small volume and \\ 
2) to transform into a final state, which consists only of two 
particles of higher energies,\\
drops extremely rapidly with rising $N$. 

Therefore, these processes drive the system towards a steady state.
Due to the broken symmetry between the rates of direct and 
inverse processes, this steady state does not coincide with a pure 
equilibrium state. 

The conditions {\bf (iii)-(iv)}, often applied for the analysis of
local equilibrium, are generally weaker than the requirement of full
local equilibrium usually imposed in the macroscopic models. At low 
energy densities, when multiparticle processes are rare, the steady 
state coincides practically with the equilibrium one. At higher energy
densities, the difference between the states becomes more and more
significant.

One characteristic feature of the steady state would be a strong
enhancement of pions, accompanied by a suppression of (many-body 
decaying) resonances. This is due to the absence of an effective 
feeding mechanism. This feature of the steady state can explain the 
fact why conventional thermal models considerably underestimate yields 
of pions at energies of $E > 10$ AGeV.
These results are typical for a large family of microscopic (cascade-, 
transport-, string-) models, which describe hadronic and nuclear 
interactions without invoking the hypothesis of quark-gluon plasma 
(QGP) creation. 

Non-equilibrium thermodynamics of irreversible processes \cite{GlPr71} 
finally comes to high energy physics, where the conservation of mass 
and particle number, conventional in statistical physics, is obviously
violated. The number of possible reaction channels is three order
of magnitude higher than in simple chemical reactions (see, however, 
the role of the equilibrium concept in biochemical/biophysical
processes). 
Therefore, it is a hopeless task to solve the rate equations
analytically. On the other hand, microscopic models for hadronic and 
nuclear collisions provide a very useful tool to probe these 
fundamental features of nature at very small space and time scale. 
The non-equilibrium aspects of heavy ion collisions are interesting 
and require further investigations.

{\bf Acknowledgements:} 
Discussions with M. Belkacem, M. Gorenstein and L. Satarov are 
thankfully acknowledged.
L.B. and E.Z. are grateful to the Institute for Theoretical Physics, 
Goethe University, Frankfurt am Main, for the warm and kind 
hospitality. This work was supported by the Graduiertenkolleg f{\"u}r 
Theoretische und Experimentelle Schwerionenphysik, BMBF, GSI, DFG,
and the A. v. Humboldt--Stiftung.

\end{document}